\documentclass[aps, pra, reprint, superscriptaddress]{revtex4-2}

\usepackage{amsmath}
\usepackage{amssymb}
\usepackage{graphicx}
\usepackage{hyperref}
\usepackage{siunitx}
\usepackage{xcolor}
\usepackage{changes}
\usepackage{comment}
\usepackage{siunitx}
\usepackage{bm}

\allowdisplaybreaks

\begin{document}

\title{Ultracold Sticky Collisions: Theoretical and Experimental Status}

\author{Roman Bause}
\author{Arthur Christianen}
\author{Andreas Schindewolf}
\affiliation{Max-Planck-Institut f\"{u}r Quantenoptik, 85748 Garching, Germany}
\affiliation{Munich Center for Quantum Science and Technology, 80799 M\"{u}nchen, Germany}
\author{Immanuel~Bloch}
\affiliation{Max-Planck-Institut f\"{u}r Quantenoptik, 85748 Garching, Germany}
\affiliation{Munich Center for Quantum Science and Technology, 80799 M\"{u}nchen, Germany}
\affiliation{Fakult\"{a}t f\"{u}r Physik, Ludwig-Maximilians-Universit\"{a}t, 80799 M\"{u}nchen, Germany}
\author{Xin-Yu~Luo} \email{Corresponding author: \\E-Mail: xinyu.luo@mpq.mpg.de}
\affiliation{Max-Planck-Institut f\"{u}r Quantenoptik, 85748 Garching, Germany}
\affiliation{Munich Center for Quantum Science and Technology, 80799 M\"{u}nchen, Germany}

\date{\today}

\begin{abstract}
Collisional complexes, which are formed as intermediate states in molecular collisions, are typically short-lived and decay within picoseconds. However, in ultracold collisions involving bialkali molecules, complexes can live for milliseconds, completely changing the collision dynamics. This can lead to unexpected two-body loss in samples of nonreactive molecules. During the last decade, such ``sticky'' collisons have been a major hindrance in the preparation of dense and stable molecular samples, especially in the quantum-degenerate regime. Currently, the behavior of the complexes is not fully understood. For example, in some cases their lifetime has been measured to be many orders of magnitude longer than recent models predict. 
This is not only an intriguing problem in itself but also practically relevant, since understanding molecular complexes may help to mitigate their detrimental effects. Here, we review the recent experimental and theoretical progress in this field. We treat the case of molecule-molecule as well as molecule-atom collisions.

\end{abstract}

\maketitle

\section{Introduction}
Intermediate complexes play a significant role in collisions and reactions between molecules~\cite{Atkins_1994, Zewail_2000}. A description of their formation and behavior is therefore crucial for understanding molecular scattering at the quantum level, a goal which is at the forefront of molecular physics and chemistry~\cite{Heazlewood_2021, Toscano_2020}. A particularly interesting example is that of the so-called sticky collisions of ultracold ground-state bialkali molecules, which have puzzled researchers for years. 

The ``sticky mystery'' started when multiple experiments~\cite{Takekoshi_2014,Park_2015,Guo_2018,Ye_2018,Jag-Lauber_2018} unexpectedly observed loss in samples of chemically nonreactive bialkalis~\cite{Zuchowski_2010} in the internal ground state. The loss had the character of a two-body process. In most cases, it was close to universal, which means that effectively every collision leads to loss of both of the particles involved~\cite{Idziaszek_2010}. If the molecules are nonreactive and the collision energy is small, this should be impossible, since energy and momentum conservation forbid any changes to the chemical bonds or internal states of the involved particles. An initial proposed explanation was that, upon collision, the molecules form extremely long-lived complexes~\cite{Mayle_2012, Mayle_2013}. This hypothesis was supported by Hu \textit{et al.~}\cite{Hu_2019}, who directly detected the complexes. 

However, it remained unclear what causes the complexes to be lost. In 2020, two groundbreaking experiments showed that they could effectively be lost after being excited by light from optical dipole traps~\cite{Liu_2020, Gregory_2020}. Because these results matched theoretical predictions~\cite{Christianen_2019a}, it was then believed that sticky collisions were basically understood. However, three other experiments soon showed that, for certain other bialkali molecules, the loss persisted in the absence of trapping light~\cite{Bause_2021,Gersema_2021}, proving that the understanding was still incomplete.

From an experimental perspective, sticky collisions are problematic because they limit the lifetime and density of ultracold samples. They also hinder evaporative cooling, a staple technique used to reach quantum degeneracy~\cite{Valtolina_2020, Li_2021, Schindewolf_2022} and make it hard to introduce scattering resonances~\cite{Yang_2019,Wang_2021,Son_2022,Park_2022,Chen_2022}, which would greatly enhance the ability to control molecular interactions. Solving these problems would enable crucial new applications of ultracold molecules, such as quantum simulation of systems with strong dipolar interactions~\cite{Astrakharchik_2007, Matveeva_2012, Peter_2012, Baranov_2012, Zeng_2014, Dutta_2015, Wu_2016, Schmidt_2021}. 

From a theory standpoint, the complexes also pose an interesting challenge. Despite the fact that they only consist of four atoms, they are beyond the level of complexity that can be handled in state-of-the-art quantum dynamics calculations. In addition, existing effective models are insufficient: though their results have been experimentally confirmed in some cases~\cite{Liu_2020, Gregory_2020, Gregory_2021}, in others they disagree with experimental results by up to five orders of magnitude~\cite{Bause_2021, Gersema_2021, Nichols_2022}.

Recent reviews have treated the creation of ultracold molecules and the general understanding of cold collisions~\cite{Quemener_2012, Balakrishnan_2016, Quemener_2018, Liu_2021a, Liu_2021b, Zhao_2022}. Here, we focus specifically on sticky collisions. We give an overview of recent experimental and theoretical results, and discuss possible solutions for the large disagreement between them.

In Section \ref{sec:theory1}, we explain the phenomenon of sticky collisions and introduce the accepted method for predicting complex lifetime. In Section \ref{sec:experiment}, we describe the existing experimental methods to probe sticky collisions. We give an overview of previous results and their agreement (or lack of agreement) with theory. Section \ref{sec:theory2} contains a discussion of possible effects which were not considered in previous models. Section \ref{sec:atom-diatom} focuses on the related problem of sticky collisions between atoms and diatoms, in the hope that a combined approach will lead to better understanding of both. Finally, in Section \ref{sec:future}, we suggest research directions, both experimental and theoretical, which we believe to be promising in order to resolve the previously discussed discrepancies.

\section{Established theory framework}
\label{sec:theory1}

\subsection{Complex lifetime from RRKM theory}
\label{sec:rrkm}

\begin{figure}[tbp]
    \centering
    \includegraphics[width=3.33in]{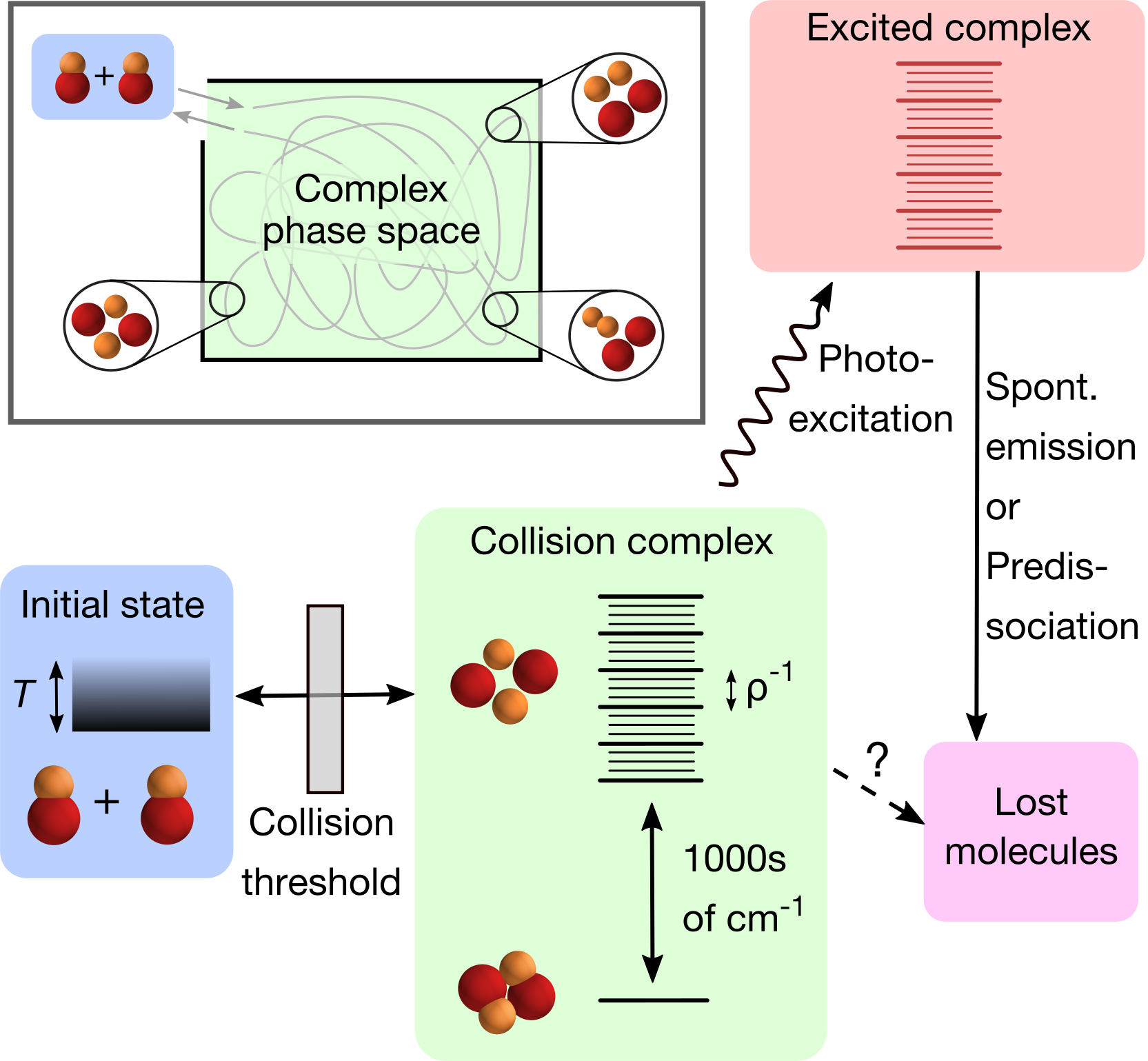}
    \caption{Schematic diagram of sticky collisions of nonreactive molecules.
    The box on the top left shows a classical picture. When molecules collide and enter the complex phase space, the path they follow is assumed to be random. The rest of the figure shows the possible pathways of sticky collisions. The kinetic-energy distribution of the initial ground-state molecules is set by the temperature. When they collide, they need to pass the threshold before they enter the short-range potential and form a complex. The number of available states in the complex may be increased depending on which quantum numbers are conserved (illustrated by additional smaller lines). The complex can be electronically excited by photons from the trapping laser and subsequently decay to other states. Otherwise, it can dissociate into ground-state molecules, or another hypothetical loss mechanism may occur.}
    \label{fig:diag_stick}
\end{figure}

The concept of sticky collisions was introduced in the works by Mayle \emph{et al.\ }\cite{Mayle_2012, Mayle_2013}. The main idea is illustrated in Fig.\ \ref{fig:diag_stick}. When two bialkali molecules collide, they enter the large phase space of the collision complex. Due to their strong and anisotropic interactions, the molecules move chaotically through the available configurations \cite{Mayle_2012, Mayle_2013,Croft_2014,Croft_2017a,Croft_2017b} before eventually dissociating when they by chance go back to their respective initial states. The ``sticking time" is in this picture naturally related to the size of the available phase space divided by the size of the ``opening" through which the particles can leave.

These ideas are more precisely formulated in a quantum-mechanical way in the statistical Rice-Ramsperger-Kassel-Marcus (RRKM) theory~\cite{Rice_1927, Kassel_1928, Marcus_1952}. In this formalism the lifetime of the complex, $\tau_{\mathrm{RRKM}}$, follows from a simple relation between the density of states $\rho$ and the number of outgoing states $N_{\mathrm{out}}$~\cite{Mayle_2012}:
\begin{equation} \label{eq:RRKM}
    \tau_{\mathrm{RKKM}}=\frac{2 \pi \hbar \rho}{N_{\mathrm{out}}}.
\end{equation}

Quantum mechanically, the notion of chaos implies that (almost) every eigenstate is delocalized over the entire accessible phase space~\cite{DAlessio_2016}. When these eigenstates couple into the scattering continuum, they give rise to scattering resonances which in turn lead to sticky collisions. Since the RRKM model is statistical, one needs to average over multiple resonances for the model to be predictive.

Applying this model to ultracold bialkali molecules yields sticking times which are very long compared to molecular timescales. This is because $N_{\mathrm{out}}$ is small and $\rho$ is very large. The reason that $N_{\mathrm{out}}$ is small is simple: If molecules are in their absolute ground state and are sufficiently cold, no inelastic collision channels are energetically accessible, thus $N_{\mathrm{out}}=1$. For reactive collisions $N_{\mathrm{out}}$ is larger, but for bialkali molecules the value is typically still modest. For example, for KRb+KRb it is approximately 100~\cite{Liu_2020}. The available phase space is very large because of the strong chemical interaction. For example, the interaction energy of two NaK molecules is $\sim \SI{4500}{cm}^{-1}$, and the rotational and vibrational constants are $\SI{0.095}{cm}^{-1}$ and $\SI{124}{cm}^{-1}$. If the interaction energy is turned into kinetic energy, up to 200 rotational levels and 35 vibrational levels can be occupied (assuming a harmonic vibrational potential). Taking into account the degeneracies of the rotations, this leads to $\sim$20000 rotational states for a single molecule. Since the degrees of freedom of two molecules plus their relative motion need to be considered, the number of involved states becomes huge. It can grow even orders of magnitude further in the presence of external fields breaking angular momentum conservation or when the hyperfine state can change during the collision.

The large number of quantum states is exactly the reason why one needs to resort to statistical models in the first place, as it makes rigorous quantum mechanical calculations computationally intractable. Some of the most computationally demanding scattering calculations to date were carried out for alkali atom-diatom collisions \cite{Croft_2017a,Croft_2017b}, which have three fewer motional degrees of freedom. This shows that moving to a full quantum description of molecule-molecule bialkali collisions will probably not be possible in at least the next decade, especially for the heavier species, for which $\rho$ is generally larger. Even if these calculations were tractable, they are sensitive to the details of the interaction potential, which is hard to determine with sufficient precision. This means that only qualitative conclusions could be drawn from this kind of calculation.

Exactly computing the density of states is equally difficult as solving the scattering problem. However, here one can resort to quasiclassical approximations, which are believed to be accurate because the involved rovibrational quantum numbers are high. In Ref.~\cite{Christianen_2019b} such a quasiclassical method was proposed to estimate $\rho$ for realistic interaction potentials~\cite{Christianen_2019a}. In the absence of external fields and when the nuclear spin states of the atoms are conserved, Eq.~\ref{eq:RRKM} typically yields $\tau_{\mathrm{RKKM}}$ ranging from few to hundreds of microseconds for collisions between bialkali molecules.

Note that this discussion is specific to bialkali dimers. For example, many other molecules have much weaker interactions and would therefore not be expected to be sticky. This especially holds for typical chemically stable molecules naturally occurring in the gas phase. Even the triplet LiNa molecule~\cite{Rvachov_2017} is in this sense different from the other investigated bialkalis because it is very light and the interaction potential is comparatively shallow. Even in the case of very strong interactions, sticky collisions are not guaranteed. For example, CaF is highly reactive~\cite{Sardar_2022}, leading to a large value of $N_{\mathrm{out}}$.

\subsection{Complex-mediated loss via photoexcitation} \label{sec:Photoexcitation}

The molecules sticking together for a long time when they collide does however not explain why they are lost. Several processes have been suggested, such as collisions with a third molecule \cite{Mayle_2013} and excitation by the trapping laser \cite{Christianen_2019c} (see Fig.~\ref{fig:diag_stick}).
In typical experiments, where molecules are held in optical dipole traps, the latter mechanism is predicted to be strong enough to excite every complex which is formed. 

Why can the trapping laser electronically excite the collision complexes? For the free molecules the trapping laser is detuned from any electronic transition so that the molecules cannot be photoexcited. However, as the geometry of the complex is continuously evolving, the gap between its electronic ground and excited states changes, sometimes matching the trap laser frequency. Again one can take a classical approximation and assume that the geometry of the complex does not change upon electronic excitation. This way, the photoexcitation rate can be computed statistically as an average over phase space~\cite{Christianen_2019c}. At typical intensities, the resulting rate is one to two orders of magnitude higher than the expected dissociation rate of the complexes~\cite{Christianen_2019c}. Once a complex has been excited, it can decay in multiple ways, such as spontaneous emission or dissociation into various asymptotic states. This makes it very unlikely that the colliding molecules go back to their original states, causing them to be effectively lost.

Given this mechanism, the rate equations for the molecule and complex populations ($n$ and $n_c$) take the form
\begin{align}
    \dot{n}&=-k_2 n^2 +\frac{2 C^{-2} n_c}{\tau_{\mathrm{RRKM}}}, \\
    \dot{n}_c&=\frac{k_2 n^2}{2} -\frac{ C^{-2} n_c}{\tau_{\mathrm{RRKM}}} -k_I I n_c.
\end{align}
Here $k_2$ is the molecular scattering rate coefficient, $k_I$ is the laser excitation rate coefficient, $I$ is the laser intensity and $C^{-2}$ is a factor which originates from quantum defect theory and describes the probability to cross the long-range part of the potential \cite{Idziaszek_2010,Christianen_2021,Croft_2021}. This factor $C^{-2}$ \cite{Christianen_2021} is implicitly also included in $k_2$ and will be described in more detail in Sec.\ \ref{sec:theory2}. To accurately describe experimental data, it is often necessary to also model effects resulting from inhomogeneous density distribution, one-body loss, and evaporation~\cite{Bause_2021,Voges_2021}.

Other loss processes, such as three-body collisions, non-excited complexes falling out of the trap, spontaneous decay into stable four-atom states, or excitation by black-body radiation, seem to be orders of magnitude too slow compared to $\tau_{\mathrm{RRKM}}$~\cite{Christianen_2019b,Man_2022}. Therefore, trapping ground-state molecules in the absence of intense laser light should lead to a suppression of the loss. 
\section{Experimental verification}
\label{sec:experiment}
\subsection{Creation and detection of ultracold molecules}
Ultracold diatomic molecules in the electronic and rovibrational ground state are typically created using a method established by Ni \textit{et al.\ }in 2008~\cite{Ni_2008}. To do this, a mixture of alkali atoms is prepared in an optical dipole trap and a Feshbach resonance is used to associate molecules with a typical binding energy on the order of $10^{-4}\,\text{cm}^{-1}$. The molecules are then transferred into the ground state, with a binding energy on the order of $10^{3}\,\text{cm}^{-1}$, using stimulated Raman adiabatic passage~\cite{Bergmann_1998, Vitanov_2017}. This method has been used with minor modifications for a number of molecular species~\cite{Danzl_2010, Takekoshi_2014, Molony_2014, Park_2015, Guo_2016, Rvachov_2017, Voges_2020, Bause_2021b, Leung_2021, Cairncross_2021, Stevenson_2022}. In recent years, direct laser cooling of molecules has made quick progress, enabling the cooling of species other than bialkalis into the ultracold regime~\cite{Fitch_2021}. However, signatures of sticky collisions have not been observed with directly cooled molecules yet, as the collision experiments performed in such systems used highly reactive CaF molecules~\cite{Cheuk_2020, Anderegg_2021}.

Most ultracold-molecule experiments use absorption or fluorescence imaging on atomic or molecular transitions to detect the molecules. These detection methods are typically specific to single quantum states and fail for complexes. However, if complex formation leads to the loss of the colliding molecules, this can be observed as two-body decay of the molecule number. Of course, this method is limited, as it only tells us that there is some destructive two-body process, but no details about the mechanism. Still, clear experimental indications of sticky collisions were found in this way~\cite{Takekoshi_2014, Park_2015, Guo_2018, Ye_2018, Jag-Lauber_2018, Gregory_2019, Voges_2020, He_2021}. 

Some clues can already be extracted by comparing the observed loss rate to the universal rate, although the precision is limited by difficulty of measuring the molecule density.  For $^{87}$Rb$^{133}$Cs and for $^{23}$Na$^{87}$Rb in its first vibrationally excited state, these experiments found loss coefficients below the universal value~\cite{Gregory_2019, Ye_2018}. On the other hand, for $^{23}$Na$^{40}$K, loss rates significantly above the universal value were found~\cite{Bause_2021}. The deviations from universality would imply that the short-range loss probability, or the $y$-parameter central in quantum defect theory \cite{Idziaszek_2010,Gregory_2019,Croft_2020,Christianen_2021}, is smaller than 1. In the RbCs experiment it was found to be $y=0.26(3)$~\cite{Gregory_2019}.

Two improved methods, which allow extracting more information about complex formation and loss, have subsequently been demonstrated. The first is to detect the complexes directly, the second is to modify the molecule-trapping potential and see whether this changes the two-body loss. In the following, we will describe these two methods in more detail.

\subsection{Detection of complexes}
\begin{figure}[tbp]
    \centering
    \includegraphics[width=3.33in]{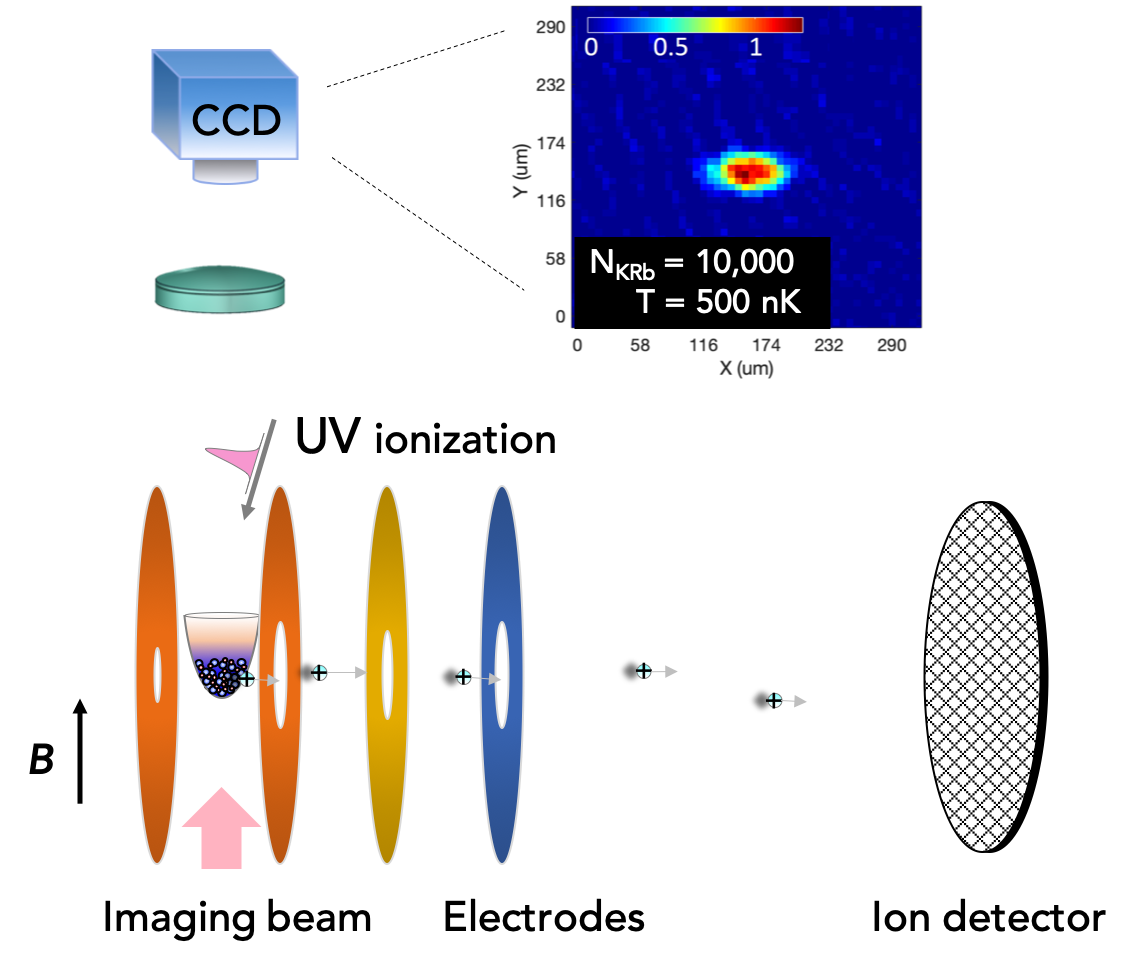}
    \caption{Schematic of the apparatus used in Refs.\ \cite{Hu_2019, Liu_2020, Liu_2020a, Hu_2020, Liu_2021, Nichols_2022}. $^{40}$K$^{87}$Rb molecules are prepared in an optical dipole trap. UV ionization and ion detection with velocity map imaging allows the detection of intermediate complexes and reaction products.}
    \label{fig:ni_lab_apparatus}
\end{figure}

An apparatus which can detect arbitrary intermediate complexes and reaction products of collisions between $^{40}$K$^{87}$Rb molecules has been constructed by the group of Kang-Kuen Ni~\cite{Hu_2019, Liu_2020a} (see Fig.\ \ref{fig:ni_lab_apparatus}). It is based on a combination of photoionization, velocity-map imaging and mass spectroscopy, similar to previous experiments on three-body recombination in Rb gases~\cite{Haerter_2013, Wolf_2017}. The molecule sample is surrounded by a hollow ultraviolet beam whose photon energy is above the ionization threshold of the complexes. Thus, the formed complexes are immediately ionized and can subsequently be imaged on a microchannel plate. Controlling the electric field on their flight path allows resolution of their mass, such that the chemical composition of the complexes can be inferred. 

\begin{figure}[tbp]
    \centering
    \includegraphics[width=3.33in]{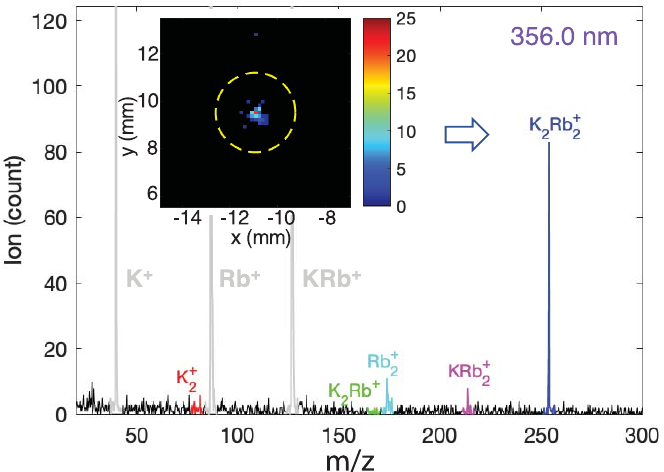}
    \caption{Direct detection of reaction products and intermediate complexes from KRb + KRb collisions with ionization and mass spectroscopy. The inset shows a velocity map of detected K$_2$Rb$_2^+$ ions. Figure adapted from Ref.\ \cite{Hu_2019}.}
    \label{fig:ni_lab_mass_spectrum}
\end{figure}

In this way, it was proved that a K$_2$Rb$_2$ complex is indeed formed in collisions~\cite{Hu_2019} (see Fig.\ \ref{fig:ni_lab_mass_spectrum}). By temporarily shutting the dipole trap off, it was also shown that the complexes indeed scatter photons from the 1064-nm dipole trap, reducing their lifetime significantly~\cite{Liu_2020}. The method has also been used to determine the distribution of rotational states created in the exothermic reaction 2KRb $\to $ K$_2$ + Rb$_2$~\cite{Hu_2020, Liu_2021}. It was determined that nuclear spin is conserved in this reaction.

Direct detection of complexes is a very useful method, as it gives a detailed picture of what goes on in a collision. However, it requires photoionization, electric fields to guide ions, and ion detection, in addition to the already complicated apparatus needed to make ultracold ground-state molecules in the first place. Currently, the Harvard experiment is the only one of this type, although the construction of a similar setup with NaK was recently started in the group of Silke Ospelkaus at the University of Hannover.

\subsection{Indirect measurement of complex properties}
\begin{figure}[tbp]
    \centering
    \includegraphics[width=3.33in]{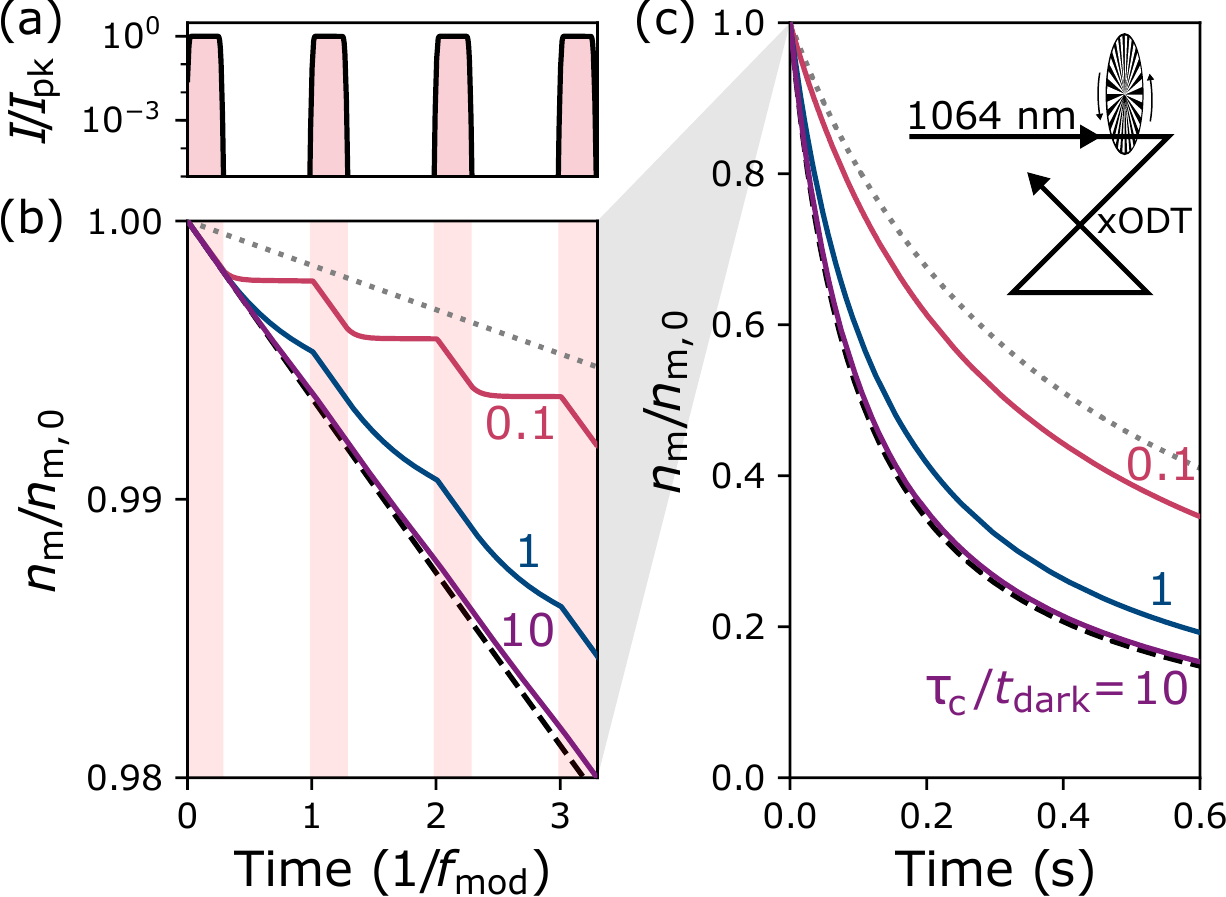}
    \caption{Principle of a chopped dipole trap. (a) Intensity over time. During the dark times of the trap, the intensity is strongly reduced. (b) Expected normalized molecule density over time in a chopped trap. The three curves show the expected behavior for ratios between sticking time and dark time of 0.1, 1, and 10. (c) Extension of (a) for long timescales. Here, the high-frequency components caused by the dipole trap switching are averaged out. Figure reproduced from Ref.\ \cite{Gregory_2020}.}
    \label{fig:chopped_trap_gregory}
\end{figure}

Assuming that the loss of complexes is dominated by photon scattering, there is another way of investigating it: significantly reduce the intensity of the dipole trap and see if this decreases the observed two-body decay rate. This has been experimentally realized in two ways. 
The first is using a chopped dipole trap, i.e., a trap whose intensity is periodically modulated in a square-wave pattern. Due to the inertia of the molecules, they experience a time-averaged potential and remain trapped if the modulation frequency is much higher than the harmonic trap frequency. During the periods where the trap is on, complexes are still likely to scatter at least one photon and are thus lost. However, if the off-times are at least comparable to the complex lifetime against dissociation into ground-state molecules, and the intensity is low enough, the time-averaged two-body loss rate should be reduced (see Fig.\ \ref{fig:chopped_trap_gregory}). Because the intensity modulation can cause loss in itself, it is difficult to directly compare lifetimes measured in modulated and non-modulated traps. This can be fixed by adding a continuous laser, a ``kill beam", to the modulated trap, and to compare the results with this kill beam on or off. By varying the kill-beam power, one can probe the necessary intensity to excite a complex during its lifetime. Indeed, a reduction of loss in such an experiment was observed by Simon Cornish's group with $^{87}$Rb$^{133}$Cs~\cite{Gregory_2020, Gregory_2021}. In contrast, no reduction was found with $^{23}$Na$^{39}$K, $^{23}$Na$^{40}$K, and $^{23}$Na$^{87}$Rb~\cite{Gersema_2021, Bause_2021}.

To avoid heating due to the trap modulation, the molecules can be trapped at permanently low light intensity. Our team at MPQ realized this by loading $^{23}$Na$^{40}$K into a repulsive, box-shaped potential with sharp edges and very low residual intensity on its inside~\cite{Bause_2021}. No dependence of two-body decay on the light intensity inside the box could be found. Notably, in this experiment the complexes were mostly lost by falling out of the trap, rather than being excited by residual light, assuming that they scatter photons at the theoretically predicted rate. This is because the complexes do not experience a repulsive potential from the trap beams and are therefore not confined. This limited the longest complex lifetimes which could be probed to a few milliseconds. 

In principle, low-light-intensity trapping could also be achieved with static electromagnetic fields~\cite{Doyle_1995, Weinstein_1998, Bethlem_2000, Hummon_2011, Reens_2017, McCarron_2018, Segev_2019, Jurgilas_2021, Park_2022a}, or microwaves~\cite{DeMille_2004, Wright_2019}. However, bialkalis in the $X^1\Sigma^+$ ground state can not be held in any static-field traps, because they experience a negative (high-field seeking) Stark shift and have a near-zero magnetic moment. Trapping them with microwaves is theoretically possible, but has never been demonstrated.

\subsection{Experimental results}
\begin{figure}[tbp]
    \centering
    \includegraphics[width=3.33in]{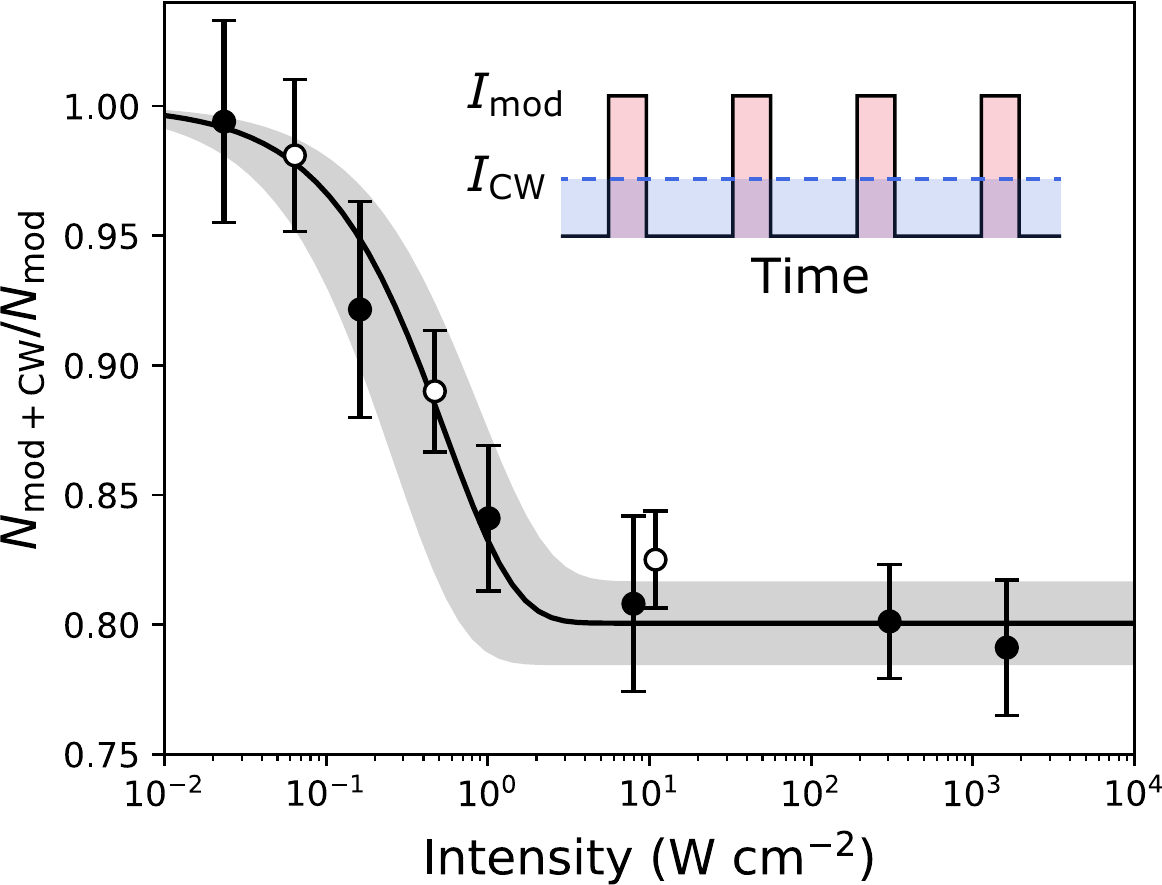}
    \caption{Observation of longer molecule lifetime in a chopped trap with $^{87}$Rb$^{133}$Cs. The ratio of remaining molecule number with (``mod+cw'') and without (``mod'') the kill beam is plotted versus its intensity. The filled and empty data points indicate measurements with the kill beam at a wavelength of 1550 and 1064 nm, respectively. The inset shows a the intensity of the chopped dipole trap (solid line) and the intensity of the kill beam (dashed line) over time. Figure adapted from Ref.\ \cite{Gregory_2020}.}
    \label{fig:chopped_trap_gregory_2}
\end{figure}

\begin{table*}[t]
    \renewcommand{\arraystretch}{1.2}
    \begin{tabular}{lllllllll}
    \hline
    Molecule & $d_0$/D & Stat. &  Nucl. spin & Setup & $\tau_{\mathrm{exp}}$ & $\tau_{\mathrm{RRKM}}$ & Comments & Refs. \\
    \hline
    $^{23}$Na$^{39}$K & 2.7 & boson & $|-3/2, -1/2\rangle$ & chopped & $> \SI{0.35}{\milli\second}$ & $\SI{6}{\micro\second}$ & Kill beam at 816, 950 and 1064 nm &~\cite{Gersema_2021}\\
    $^{23}$Na$^{40}$K & 2.7 & fermion& $|3/2, -4 \rangle \dagger$  & box & $>\SI{2.6}{\milli\second}$ &  $\SI{18}{\micro\second}$ $(\SI{4.9}{\milli\second})$ & Electric field \SI{411}{V/cm} &~\cite{Bause_2021}\\
        & & fermion & $|3/2, -4 \rangle \dagger$ & box & $>\SI{1.4}{\milli\second}$ &  $\SI{18}{\micro\second}$ $(\SI{4.9}{\milli\second})$ & &~\cite{Bause_2021}\\
        & & --- & mixed & box & $>\SI{2.3}{\milli\second}$ &  $\SI{18}{\micro\second}$ $(\SI{54}{\micro\second})$& Incoh. mixture, el. field \SI{411}{V/cm} &~\cite{Bause_2021}\\
        & & --- & mixed & box & $>\SI{133}{\micro\second}$ &  $\SI{18}{\micro\second}$ $(\SI{54}{\micro\second})$& Incoherent mixture &~\cite{Bause_2021}\\
    $^{23}$Na$^{87}$Rb & 3.2 & boson & $|3/2, 3/2\rangle * \dagger$ & chopped& $ > \SI{1.2}{\milli\second}$ & $\SI{19}{\micro\second}$ & Kill beam at 1064 and 1248 nm, &~\cite{Gersema_2021}\\
        & & & & & & & electric field $ < \SI{5}{V/cm}$ \\
    $^{40}$K$^{87}$Rb & 0.6 & fermion & $|-4,1/2\rangle$ & direct det.& $\SI{360(30)}{\nano\second}$ & $\SI{170(60)}{\nano\second}$ & Reactive, el. field \SI{17}{V/cm} &~\cite{Liu_2020}\\
    $^{87}$Rb$^{133}$Cs & 1.2 & boson & $|3/2, 7/2 \rangle * \dagger$  & chopped& $\SI{0.53(6)}{\milli\second}$ & $\SI{0.253}{\milli\second}$ & Kill beam at 1064 and 1550 nm &~\cite{Gregory_2020}  \\
        & & boson & $|3/2, 7/2 \rangle * \dagger$  & chopped & $\SI{0.8(3)}{\milli\second}$ & $\SI{0.253}{\milli\second}$ & Kill beam at 1550 nm &~\cite{Gregory_2021} \\
        & & boson & $|3/2, 5/2 \rangle $  & chopped & $2.1(1.3)\,\mathrm{ms}$ & $\SI{0.253}{\milli\second}$ & Kill beam at 1550 nm &~\cite{Gregory_2021}  \\
        & & boson & $|1/2, 7/2 \rangle$  & chopped & $>\SI{3.3}{\milli\second}$ & $\SI{0.253}{\milli\second}$ & Kill beam at 1550 nm &~\cite{Gregory_2021}  \\
    \hline
    \end{tabular}
    \caption{Experimental results for the complex lifetime in collisions between bialkalis in the electronic and rovibrational ground state. The molecular-frame dipole moment $d_0$~\cite{Gerdes_2011, Guo_2016, Ni_2010, Takekoshi_2014} is given in units of Debye. We denote the nuclear spin as $|m_{i,a}, m_{i,b}\rangle$, where $a$ and $b$ indicate the lighter and heavier atom in the molecule. Stretched hyperfine states are marked with a $*$, and ground hyperfine states with a $\dagger$. For the predicted complex lifetimes $\tau_{\mathrm{RRKM}}$ we assume no nuclear-spin changing collisions take place. The $\tau_\mathrm{RRKM}$ values in brackets take the effect of reflection from the long-range potential into account as discussed in Section \ref{sec:validity_rrkm}. The additional laser light which ensures destruction of complexes for comparison (``kill beam'') had a wavelength of 1064 nm except where stated otherwise. Where not explicitly noted, the experiments were performed at small electric and magnetic fields (below \SI{10}{V/cm} and \SI{200}{G}, respectively). For those experiments where only lower bounds of $\tau_\mathrm{exp}$ are given, these are calculated under the assumption that the complex photoexcitation rates are equal to the predicted values. The temperature of molecules in most experiments listed is between 100 and 500 nK, except for RbCs, which had a temperature of around 2 $\mu$K. For NaK, KRb and NaRb, $\tau_{\mathrm{RRKM}}$ values were computed in Refs. \cite{Christianen_2019b,Liu_2020,Liu_2022}. For RbCs, $\tau_{\mathrm{RRKM}}$ was extrapolated from the NaK result in Ref.~\cite{Christianen_2019b}. The recollision effect was studied in Ref.~\cite{Christianen_2021}. }
    \label{tab:experimental_results}
\end{table*}

At this point, experiments on sticky collisions have been performed with five different species of bialkali molecules, some in multiple hyperfine states. A list of the results is given in Table \ref{tab:experimental_results}. The first two experiments, both published in 2020, confirmed the RRKM predictions with both a reactive and a nonreactive species. With $^{40}$K$^{87}$Rb, a complex lifetime of \SI{360(30)}{ns} was found, compared to $\tau_\mathrm{RRKM} = \SI{170(60)}{ns}$. The measured photoexcitation rate was $\SI{420(90)}{Hz/(W/cm^{-2})}$, compared to the theoretically expected $\SI{400}{Hz/(W/cm^{-2})}$, however a significant nonlinear contribution was also observed~\cite{Liu_2020}. With $^{87}$Rb$^{133}$Cs, the measured complex lifetime was \SI{0.53(6)}{ms}, with $\tau_\mathrm{RRKM} = \SI{0.253}{ms}$, and the measured photoexcitation rate was $\SI{3}{kHz/(W/cm^{-2})}$ (see Fig.\ \ref{fig:chopped_trap_gregory_2}).

\begin{figure}[tbp]
    \centering
    \includegraphics[width=3.33in]{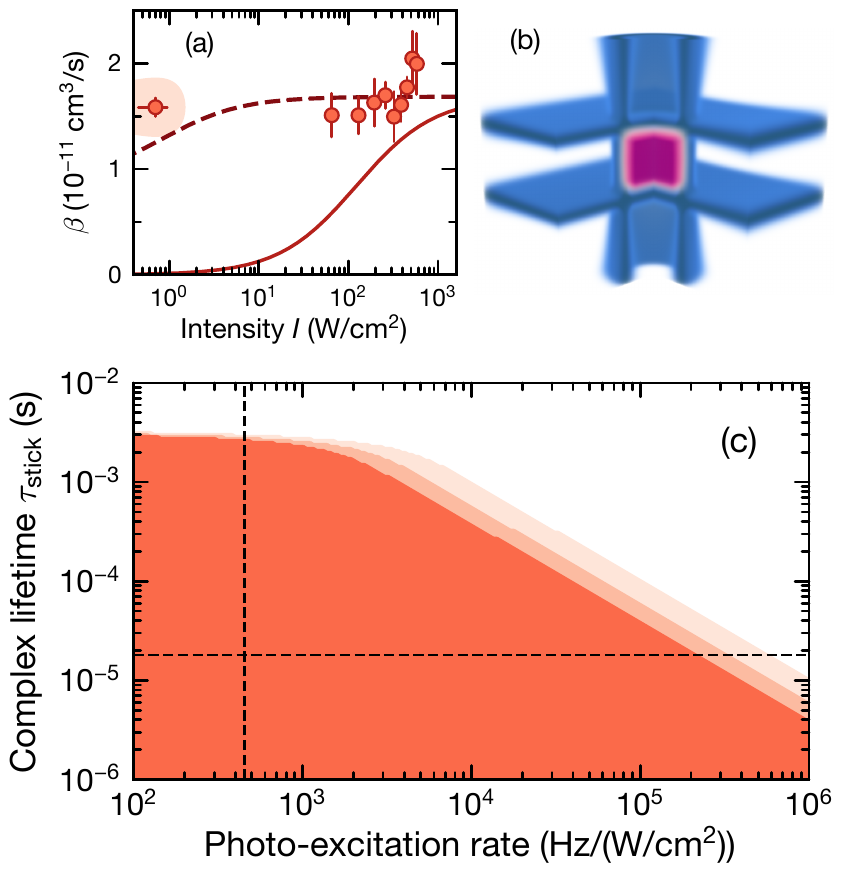}
    \caption{Complex loss of $^{23}$Na$^{40}$K in a repulsive box-shaped trap~\cite{Bause_2021}. (a)~Two-body loss coefficient $\beta$ versus 1064-nm-light intensity. The solid line indicates the predictions from Ref.\ \cite{Christianen_2019b}, the dashed line is calculated for the shortest complex lifetime which is consistent with the experimental data within $3\sigma$. (b)~Sketch of the box trap used to confine molecules at low light intensity. The molecule cloud is shown in purple. One quadrant is cut out for visibility. (c)~Parameter space excluded by the data under the assumption that complexes can only be lost by photoexcitation or by leaving the trap. The three shaded areas are excluded with $1\sigma$, $2\sigma$, and $3\sigma$ confidence, respectively. The dashed lines indicate the predictions from Ref.\ \cite{Christianen_2019b}.}
    \label{fig:box_trap_result}
\end{figure}

Following these, there were three other experiments, which used $^{23}$Na$^{39}$K, $^{23}$Na$^{40}$K, and $^{23}$Na$^{87}$Rb (see Fig.~\ref{fig:box_trap_result}). Very surprisingly, not only did these experiments disagree with the RRKM predictions, they could not reduce the loss of complexes with low light intensity at all. This was despite the fact that the predicted complex lifetimes for all these species are much shorter than that of RbCs, which should have made loss reduction even easier.
If no loss reduction is found, such as in these experiments, it is impossible to extract complex lifetimes and photoexcitation rates. Then, only a combined lower bound for lifetime and photoexcitation rate can be measured. Typically, the experimental results are presented as a lower bound to the complex lifetime given the theoretical photo-excitation rate, since this prediction is considered most reliable.

Looking at the table, there are two species where the observations agree with the RRKM predictions within a factor of two, while for three others, they disagree by at least a factor 50. There is no obvious correlation between the molecular properties or the experiment design, and the agreement with predictions. For example, one might expect that the loss rates are smaller for molecules which are in the hyperfine ground state or in a spin-stretched state (i.e., $m_F = \pm F$, and $F$ takes its maximum value), because spin-changing collisions are then forbidden. Considering $^{87}$Rb$^{133}$Cs and $^{23}$Na$^{87}$Rb in the hyperfine ground state, they should behave similarly: both are nonreactive bosons and spin-changing collisions are forbidden. Instead, in the experiments, one of them behaves as expected, the other very differently. 

Another surprising result was found at MIT, where a Feshbach resonance was observed in collisions of $^{23}$Na$^{6}$Li molecules in the $a^3\Sigma^+$ manifold \cite{Park_2022}. Although no direct measurements of complex lifetime were performed, some limits for the lifetime of the complex can be inferred from the shape of the interaction potential and the width of the resonance. It is unclear how to compare this lifetime with the sticking times measured in the other experiments, because the quasi-bound state underlying the resonance is likely of a different nature than the sticky complexes discussed before. Whether similar resonances also occur for other bialkali molecules is not obvious, since $^{23}$Na$^{6}$Li was studied in a different electronic configuration.

\section{Proposals beyond the established theory} 
\label{sec:theory2}
\subsection{The validity of RRKM theory}
\label{sec:validity_rrkm}

\subsubsection*{The assumption of chaoticity}
First we discuss the validity of the assumption  of chaotic dynamics, which is central in RRKM theory and underlying Eq.~(\ref{eq:RRKM}). Hamiltonians that exhibit quantum chaos have a special structure in their spectrum, where the energy level spacings are distributed according to a Wigner--Dyson distribution, as opposed to a regular Poisson distribution. Also the distribution of resonance widths in case of quantum chaos is known. It was shown in explicit quantum scattering calculations by Croft \emph{et al.~}\cite{Croft_2017a,Croft_2017b} that this at least holds relatively well for ultracold atom-diatom collisions of alkali atoms. 

In recent work by Man \emph{et al.~}\cite{Man_2022}, the statistical assumption for atom-diatom and diatom-diatom collisions was tested with classical trajectory simulations, and found to be valid. In order to extract the sticking times from classical trajectory simulations, it is important to set appropriate long-range boundary conditions because the classical approximation is no longer justified if the molecules move far away from each other. This explains why trajectory simulations from Ref.~\cite{Croft_2014} yielded too long sticking times, agreeing with the RRKM results from Ref.~\cite{Mayle_2012} which later turned out to be overestimated by orders of magnitude \cite{Christianen_2019b}, and why simulations from Ref.~\cite{Klos_2021} gave orders of magnitude shorter sticking times.

\subsubsection*{How predictive is the statistical model?}
Next, we discuss whether one indeed averages over many scattering resonances under experimental conditions, as required to get reliable predictions from the statistical model. As indicated in Fig.~\ref{fig:diag_stick}, the range of probed collision energies is set by the temperature, whereas the spacing of the resonances is set by the density of states $\rho$. The typical temperature is on the order of a microkelvin, and the density of states ranges from $0.005$ to $\SI{5}{\mu K^{-1}}$ for different bialkalis. Comparing these numbers, one may notice that only very few, if any, resonances are expected to be probed under experimental conditions. The original idea \cite{Mayle_2013} that many isolated narrow resonances would be included in the experimental window of collision energies, is therefore clearly not true. The applicability of the statistical model thus seems questionable.

However, if only a single or few resonances were included in the experimental window of collision energies, then why are these not resolved?  
An answer to this question was proposed in Ref.~\cite{Christianen_2021}. Here the authors use quantum defect theory (QDT) to show that a third energy scale needs to be taken into consideration: the broadening of the resonances due to the loss. If the loss rate is fast compared to the dissociation rate of the complexes, this automatically implies that the resonances are broadened much beyond their natural width. As a result, instead of averaging over many narrow isolated resonances in a certain energy window, one averages over many overlapping broad resonances at a single collision energy. If the broadening is strong enough, statistical models become valid again. In this scenario, close-to-universal loss would occur independently of the collision energy without any resonances, which agrees with the experimental observations.

In the recent work by Croft \emph{et al.}~\cite{Croft_2021} it is argued that the validity of the RRKM model is also questionable in the case where the collision has only a single open channel. Then the distribution of resonance widths is very strongly skewed towards narrow resonances, corresponding to long sticking times. This means that when averaging over a small number of resonances, one most likely finds a longer sticking time than expected from RRKM. Even when sampling over many resonances, this could still have an impact on the actual loss rate. However, the estimated increase in sticking time from Ref.\cite{Croft_2021} is not sufficient to explain the experimental results.

\subsubsection*{Threshold effects}

Another point made in Refs.\ \cite{Christianen_2021,Croft_2021} is that RRKM theory does not take into account threshold effects. For example, fermionic molecules undergoing $p$-wave collisions need to tunnel through the centrifugal barrier to reach the short-range potential. What was not considered before is that once the molecules passed the barrier, they also need to tunnel through the same barrier to leave it again, leading to longer complex lifetimes. The tunneling probability, parametrized by $C^{-2}$ in quantum defect theory, decreases with lower temperature, making the effect more pronounced for very cold samples.

The actual tunneling probabilities can be computed using QDT. For fermionic $^{23}$Na$^{40}$K in the experimental conditions of Ref.~\cite{Bause_2021}, the tunneling probability is only about $1\%$. This corresponds to the well-known fact that at ultracold temperatures $p$-wave collisions occur at much lower rates than $s$-wave collisions. It also means that complexes formed in $p$-wave collisions should be about 100 times longer-lived. This can explain the complex-lifetime measurements on identical fermionic NaK molecules. For $s$-wave collisions the factor is of order $1$, so threshold effects can not explain the observations for bosonic or distinguishable molecules. 

\subsection{Coupling to hyperfine degrees of freedom and external electric fields}

Even if RRKM theory is valid, there could still be large uncertainties in the calculation of the density of states. In the estimation of the sticking times in Ref.~\cite{Christianen_2019b} and comparison to photoexcitation rates~\cite{Christianen_2019a}, it is assumed that the motional angular momentum of the molecules is conserved during the collision. However, coupling to external electric fields or to hyperfine states of the molecules can break this angular-momentum conservation, potentially enlarging the explored phase space and the sticking time by up to four orders of magnitude \footnote{To get to four orders of magnitude, for the electric field case we assume that the projection of the angular momentum is conserved \cite{Christianen_2019b}. For the hyperfine case we assume the total angular momentum (including the nuclear spin) during the collision is conserved. In both cases we assume the entire additional phase space is explored. For the hyperfine case the precise enhancement factor depends on the size of the nuclear spins of the atoms. For example, for $^{23}$Na$^{39}$K and $^{23}$Na$^{87}$Rb it will be $\sim 10^3$, for $^{23}$Na$^{40}$K, for $^{87}$Rb$^{133}$Cs it will be $\sim 10^4$. If the sticking times are really enhanced by this additional factor, the magnetic field could also break the total angular momentum conservation, enlarging the phase space even more.}. Note that, in contrast with typical ultracold atomic collisions, the time spent in the short-range potential is much longer than in the long-range potential so that these processes most likely take place at short range.

Using classical trajectory simulations, Man \emph{et al.} \cite{Man_2022} estimated that electric fields on the order of 10 V/cm are already sufficient to break angular-momentum conservation. Indeed, these are the field strengths where the typical molecular Stark couplings are comparable to the level spacing of the complex. Such electric fields are larger than the stray fields typically present in experiments, but at least one order of magnitude smaller than those needed to polarize the molecules. In Ref.~\cite{Quemener_2022} it was shown that an external electric field can also affect the long-range part of the collision, since it brings the molecules into a superposition of multiple angular-momentum states, leading to an increased density of states. However, this effect is weaker than the short-range effect predicted in Ref.~\cite{Man_2022}, and it requires one to two orders of magnitude higher field strengths to become important.

Nuclear-spin changes during the collisions can have two important and distinguishable effects: if the molecules start in their hyperfine ground state, nuclear-spin changes can lead to an increased density of states. If the molecules start in excited hyperfine states which are not stretched, this opens up more exit channels and the additional loss pathway of hyperfine-changing collisions. These effects are strongly related, because both of them require sufficiently strong couplings between the nuclear spins and other degrees of freedom.

To estimate at which strength the hyperfine couplings will start to play a role, one can compare the inverse of the coupling strength to $\tau_{\text{RRKM}}$. The strongest effect is the coupling of the dynamically changing electric-field gradient to the nuclear quadrupole moments~\cite{Man_2022,Jachymski_2021}. Using the couplings of the free molecules, Man \emph{et al.~}\cite{Man_2022} estimated that nuclear-spin flips are unlikely to happen. However, Jachymski \emph{et al.~}\cite{Jachymski_2021} showed that the couplings are strongly geometry-dependent, and therefore nuclear-spin-changing collisions can not be ruled out. 

The hyperfine structure certainly plays a role in the case of RbCs, where hyperfine-dependent loss rates were observed in the chopped dipole trap~\cite{Gregory_2021}. This means that either the hyperfine state has a strong impact on the sticking time, or hyperfine-changing collisions lead to loss. In contrast, nuclear-spin-changing reactions do not occur for KRb~\cite{Hu_2020}. However, here the collision takes only hundreds of nanoseconds, so these results are insufficient to determine the role of nuclear-spin flips in much longer sticky collisions. For triplet molecules or atom-molecule collisions, unpaired electron spins also play a role. The couplings between the electronic and nuclear spins and the electronic spin-rotation couplings are orders of magnitude stronger than the couplings of only the nuclear spins. Therefore unpaired electron spins can change and mediate hyperfine changes during the collision.

In conclusion, it seems clear that nuclear-spin changes have some effect on the collision dynamics at least for heavy bialkalis. However, to explain the increased sticking times for the lighter bialkali molecules, the effect of the hyperfine degrees of freedom would have to be much stronger than it is for heavy species. This is the opposite of what one should expect, since both the hyperfine couplings and $\tau_{\mathrm{RRKM}}$ are smaller for lighter bialkalis. Hence, if nuclear-spin changes in the complex are responsible for longer sticking times, there must be some yet unknown ingredient to bring the predictions into quantitative agreement with the experimental observations.

\subsection{Loss processes}
\label{sec:loss_processes}
In addition to the sticking time, the complex loss rate is also of crucial importance to assess whether the molecules will survive a collision. As explained in Section \ref{sec:Photoexcitation}, the computation of the photoinduced loss rate, just like RRKM theory, is based on a quasiclassical statistical model. The validity of this model has not yet been explicitly tested. In a quantum mechanical picture, in the case of quantum chaos, all eigenstates are delocalized throughout phase space in both the ground and excited state. Therefore, every state has a small but usually nonzero Franck--Condon overlap with all the states in the excited potential. If the width of the electronic transitions is then larger than the rovibrational spacing, this means that there are no photoexcitation resonances and the complex can be excited by a continuum of laser frequencies. Since the linewidth of the typical trapping lasers is very small, the observed linewidth of the transition is limited by the lifetime of the electronic state.

Let us briefly consider the loss pathways from this electronically excited state. The simplest possibility is spontaneous emission back to some rovibrational level in the electronic ground state. Because this would be similar to an electronic transition of one of the constituent diatoms, the rate of such a process can be assumed to be on the order of \SI{10}{MHz}~\cite{Jones_2006}. Other processes include dissociation into a trimer and a free atom, or into a singlet and a triplet dimer via spin-orbit coupling. How fast these processes are depends on the details of the interaction potential, and the specific state where the complex ends up after photoexcitation. A more detailed theoretical investigation of the complex dynamics in the excited state would definitely be interesting. This could be experimentally relevant to know in which scenarios one would expect to find resolved photoexcitation lines as opposed to a continuum. For most diatom-diatom collisions, spontaneous emission would already be fast enough to lead to a continuum of excited states, but for atom-diatom collisions individual lines would be observable if this was the only or the fastest decay process.

One can also imagine loss processes other than photoexcitation playing a role. The original proposal was that this could be three-body loss \cite{Mayle_2013}, but in Ref.\ \cite{Christianen_2019b} the three-body rate was estimated to be too small. Furthermore, the complexes might escape from optical dipole traps as they are not trapped by the trapping laser. This is especially likely for repulsive traps such as the one used in Ref.\ \cite{Bause_2021}. For this process to be relevant, the sticking time typically needs to be on the scale of milliseconds. Estimation of rovibrational transition rates of non-excited complexes due to spontaneous emission and black-body coupling are given in Ref.~\cite{Man_2022}. Both processes are far too slow to occur in the $\tau_{\mathrm{RKKM}}$ timescale and require sticking times of approximately tens of seconds.
 
Another potential explanation for the loss is the presence of special features in the potentials such as conical intersections. Conical intersections are known to occur and have an influence on the collision dynamics for alkali atom-diatom collisions~\cite{Kendrick_2021}. We do not expect that such features destroy the validity of RRKM models, but they might enhance the probability of nuclear-spin flips or lead to the population of electronically excited states. In Ref.~\cite{Christianen_2019c} it was estimated that for NaK-NaK there are conical intersections close to, but still outside, the classically allowed region. However, this might need to be revisited with more accurate methods to give a clearer judgement. Then one would need to find and characterize the loss processes that could happen around such a conical intersection. What makes this explanation attractive is that the presence of a conical intersection could strongly depend on the species, and therefore explain the seemingly arbitrary differences between the bialkalis.
 
\subsection{Preliminary conclusion}
Currently, the most likely explanation of the sticky-collision mystery seems as follows. In presence of laser light, photoinduced loss is probably the dominant mechanism. To explain the experimental results for the lighter bialkalis in absence of laser light, the sticking time must either be orders of magnitude longer than predicted, or there are loss processes that are much faster than expected and can occur even during short sticking times. The most important open question from the theory side is how the nuclear-spin degrees of freedom are involved in the collision. There are some clues that understanding this will allow better predictions of the intricate behavior of complexes.

Alternatively, it is possible that the experimental results can be explained by a combination of threshold effects and external electric fields. Specifically, for the case of the hyperfine-pure samples of fermionic $^{23}$Na$^{40}$K, the $p$-wave barrier might be the cause of the much longer sticking times. If the electric fields in the experiments with (bosonic) $^{23}$Na$^{39}$K and $^{23}$Na$^{87}$Rb were large enough to break angular-momentum conservation, this problem would be solved too. To confirm this hypothesis, the sticking time should be measured as a function of electric field strength.
 
\section{Atom-diatom collisions}
\label{sec:atom-diatom}
\subsection{Differences and similarities to diatom-diatom collisions}
With the difficulties in understanding collisions between two molecules, it seems logical to first study a simpler case: the collision between a molecule and an atom. This was originally motivated by sympathetic cooling, where molecules are cooled by elastic collisions with atoms, which in turn can be evaporatively cooled~\cite{Son_2020, Jurgilas_2021}. Atom-diatom collisions are in many ways similar to diatom-diatom collisions, making them an interesting system to study to gain understanding of sticky collisions.

From the theoretical side, a three-atom system is much more tractable than a four-atom one. Indeed, full quantum-mechanical calculations of collision times have been performed for certain species~\cite{Croft_2017a, Croft_2017b}. These results can not quantitatively be compared to experiments, since it is very challenging to include spin degrees of freedom or external fields in the calculations. Furthermore, getting to experimental accuracy would require extremely accurate interaction potentials. However, for qualitative studies these calculations are valuable, for example to confirm the validity of RRKM theory~\cite{Croft_2017b}. An overview of the RRKM predictions on atom-diatom collisions can be found in Ref.~\cite{Frye_2021}. 

\subsection{Experimental results}
The typical way to experimentally study ultracold atom-diatom collisions is to associate molecules from an atomic mixture as usual, keeping some unassociated atoms in the trap afterwards. This limits the available partner atoms to those which are already part of the molecule. For reactive combinations, such as $^{23}$Na$^{39}$K + $^{23}$Na, and $^{87}$Rb$^{133}$Cs + $^{87}$Rb, it has consistently been found that the two-body rate coefficients are close to the universal limit, as expected~\cite{Voges_2021, Gregory_2021}. However, in the nonreactive case the results are much more varied. Here, we give a list of experiments done with such nonreactive combinations:

\begin{itemize}
    \item $^{23}$Na$^{6}$Li + $^{23}$Na, with the diatom not in the electronic ground state, but rather in the rovibrational ground state of the $a^3\Sigma^+$ potential~\cite{Son_2020, Son_2022}. Here, the two-body rate coefficients are strongly dependent on the hyperfine state of the atom: while the coefficient is two orders of magnitude below the universal limit for Na in $F=2, m_F=2$, it is close to universal for $F=1, m_F=1$. This suggests that the collisions do not cause electronic relaxation of the diatom, but can cause spin exchange with the atom. With this combination, the loss rates can be small enough to realize efficient sympathetic cooling~\cite{Son_2020}. In addition, a magnetic Feshbach resonance has been found at a field of \SI{978}{G}~\cite{Son_2022}. 
    \item $^{23}$Na$^{39}$K + $^{39}$K~\cite{Voges_2021}. Again, the collision rate coefficients are strongly dependent on the hyperfine state of the atomic collision partner. In the most favourable case of K in $F=1, m_F=-1$, the rate is four orders of magnitude below the universal limit. In the spin-stretched state (K in $F=1, m_F=1$) it is only one order of magnitude below the universal limit.
    \item $^{23}$Na$^{40}$K + $^{40}$K~\cite{Yang_2019, Wang_2021, Su_2022}. In this case, the dependence of the observed two-body rate coefficients on the atomic hyperfine state is much weaker than with bosonic potassium, with less than one order of magnitude difference between the different states. The smallest observed coefficient is about ten times below the universal value. Some of the investigated channels exhibit Feshbach resonances, one of which has been used to associate triatomic molecules~\cite{Yang_2022, Yang_2022a}.
    \item $^{40}$K$^{87}$Rb + $^{87}$Rb~\cite{Nichols_2022}. Here, only one hyperfine channel was investigated, with the diatom in $m_{i,\mathrm{K}} = -4, m_{i,\mathrm{Rb}} = 1/2$ and the atom in $F=1, m_F=1$. In this case, the rate is close to universal. The formation of three-body complexes and their excitation by trap photons was directly observed. Notably, the measured complex lifetime of \SI{0.39(6)}{\milli\second} is five orders of magnitude larger than the RRKM prediction.
    \item $^{87}$Rb$^{133}$Cs + $^{133}$Cs~\cite{Gregory_2021}. For the case where both the diatom and the atom are in the hyperfine ground state, the rate coefficient is within a factor of three of the universal rate. The light-intensity dependence of the molecule loss rate was also probed via the chopped-trap method, but no effect of intensity could be seen.
\end{itemize}

\subsection{Interpretation of experimental results}
Together, these experiments give an inconclusive picture. We can see that the measured two-body rate coefficients are near-universal in some cases, but multiple orders of magnitude smaller in others, with a strong dependence on the atomic spin. This is not surprising in itself, because the atom brings an unpaired electron into the collision, whose spin can couple strongly to both hyperfine and rotational degrees of freedom of the diatom. If the scattering is non-universal, the collision rate is determined by the scattering length, which can depend strongly on many parameters. The smaller the mass of the collision partners, the more likely the loss is to be non-universal due to the smaller density of states. This explains the strong loss reduction found in some light alkali systems.

Another idea, brought up in Ref.\ \cite{Voges_2021}, is that atom-diatom complexes may exhibit resolved photoexcitation lines. For diatom-diatom complexes such lines are unlikely to be observable (see Sec.\ \ref{sec:loss_processes}), but since the density of states of the excited atom-diatom complexes is much lower than for the diatom-diatom case, this might be different here. If such lines were near the optical trap wavelength, this would be another explanation for the dependence of loss rate coefficients on the atomic hyperfine state. This could be experimentally tested by scanning the trapping-laser wavelength.

There is evidence that the Feshbach resonances observed for $^{23}$Na$^{40}$K + $^{40}$K result from coupling to long-range states, in which the rovibrational structure of the diatom remains almost unchanged~\cite{Wang_2021}. This is in stark contrast to short-range complex states with their chaotic rovibrational dynamics. In the spectrum of the Hamiltonian there are many more rovibrationally excited states than long-range bound states. These long-range bound states exist for different hyperfine channels. Since the resonances result from coupling between these different hyperfine states, they can be tuned with the magnetic field. Furthermore, the collision energy is close to the threshold, and this is exactly the energy range in which this type of long-range bound states can be found. That is why, especially for atom-diatom collisions, it is not unlikely to find resonances due to these long-range bound states.

\begin{figure}[tbp]
    \centering
    \includegraphics[width=3.33in]{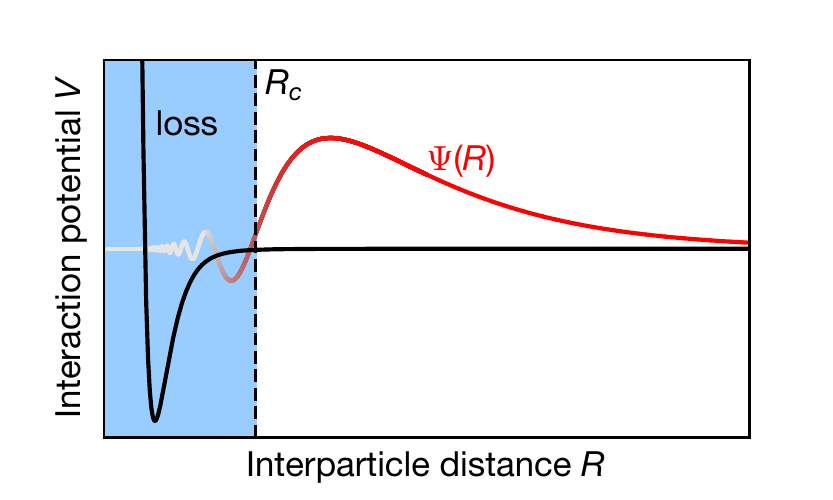}
    \caption{Schematic wave function $\Psi(R)$ (grey-red) of a bound state leading to a long-range resonance. In black the interaction potential $V$ is plotted as a function of the interparticle distance $R$. The collision threshold for the short-range loss, which happens inside the blue region, is indicated by $R_c$.}
    \label{fig:wavefc_longrange}
\end{figure}

At the scattering resonances the loss is larger than universal, even in presence of a background with significant loss. This might seem counterintuitive because at the universal rate, all collisions should already lead to loss. However, more precisely, this includes only collisions where the molecules reach the short-range regime. In the case of long-range bound states, the wave function extends beyond the region of short-range loss as illustrated in Fig.~\ref{fig:wavefc_longrange}. In this case the collision partners are temporarily trapped in the long-range potential, leading to an enhanced rate of collisions reaching the short range. This can lead to coherently enhanced loss rates even in presence of strong background loss.

For now, it appears that atom-diatom collisions exhibit much more structure than diatom-diatom ones, with observable dependence on internal states, fields, etc. However, there are still significant shortcomings to our understanding of these collisions, and further investigation will certainly be needed. This path may lead to improved sympathetic cooling of atom-molecule mixtures and efficient association of polyatomic molecules.

\section{Future perspectives}
\label{sec:future}
Further experimental and theoretical work is going to be needed to find an explanation for the sticky-collision mystery. From the experimental side, it will be necessary to gather data from more species and under a wider range of conditions. We believe that the following experiments can be realistically done in the next few years to advance our understanding. 

Firstly, chopped-trap or box-trap experiments can be performed on further biakali species, or in different hyperfine states. The most readily available candidates for new species are $^{23}$Na$^{133}$Cs, and $^{133}$Cs$_2$, but also others, such as $^{40}$K$^{133}$Cs and $^{6}$Li$^{40}$K, may soon become available~\cite{Park_2022, Stevenson_2022, Jag-Lauber_2018, Groebner_2016, Yang_2020}. The electric field should be precisely controlled to ensure that its influence on angular-momentum conservation can be understood. In addition, we hope that there will soon be new setups for direct detection of complexes formed in collisions between nonreactive molecules, despite the complexity of this approach. This may also allow detection of nuclear-spin changing collisions. Another avenue is to study ultracold collisions of non-bialkali molecules. For example, ground-state $^{88}$Sr$_2$ was recently created for the first time~\cite{Leung_2021}. Experiments on these species may help reveal the influence of the electronic structure on the collision dynamics.

Second, valuable information can be gained from atom-molecule collisions. For example, it will be interesting to see a direct comparison between the complex lifetimes in reactive and nonreactive collisions involving the same molecule, such as KRb + K and KRb + Rb. Since they are very different in these cases, comparing them may help with understanding the timescale of nuclear-spin-changing collisions. Probing more different spin-state combinations may also yield interesting results. Going even further in this direction, comparing Sr$_2$ + Sr collisions with different isotopologues could be used to disentangle the effects of nuclear spin, since $^{86}$Sr and $^{88}$Sr have a nuclear spin of zero, but $^{87}$Sr has a nuclear spin of 9/2~\cite{Jachymski_2021}. Another interesting subject of study is the systematic comparison between the isotopologues of NaK + K, at different magnetic fields and in different spin states. If Feshbach resonances are found for $^{23}$Na$^{39}$K + $^{39}$K, similar to the case with $^{40}$K, this could shed light on the relation between the properties of these resonances and the value of the background loss rate. Finally, it would be interesting to probe the laser-frequency dependence of atom-diatom-complex loss to determine if there are resolvable photoexcitation lines.

From the theory side, for the molecule-molecule collisions, open research directions include a more detailed study of hyperfine-changing collisions. In particular, a combination of methods from Refs.~\cite{Man_2022, Jachymski_2021} seems promising. Furthermore, a systematic search of the potential energy surfaces for special features like conical intersections, such as partially performed in \cite{Christianen_2019a}, could include or exclude this as a possible explanation. In the long run, although a full quantum model seems to be unrealistic for molecule-molecule collisions, we hope a universal effective model can be developed to predict the behavior of complexes for a wide range of molecules.

It would also be worthwhile to develop a more complete description of the photoexcitation process. Both to test the statistical model and to find out what happens to the complex after the excitation. The final products could be experimentally observable. Furthermore, the linewidths of the electronic transitions could be relevant for atom-molecule collisions, to see if individual lines could be resolved. 

For atom-molecule collisions full rovibrational quantum scattering calculations have been carried out already, giving insight into the chaoticity of the dynamics. Further research in this direction, especially including the spin degrees of freedom, which might be possible for light systems, would be of great importance.

The sticky-collision mystery has been around for years and many attempts have been made to solve it. What unites all these attempts is that, so far, they have succeeded only in making the problem even more mysterious. However, with a comprehensive experimental and theoretical effort now underway, we are certain that this will not be the case much longer.

\section{Acknowledgments}
This work was initially prompted by stimulating discussions at the online workshop ``Understand and control collisions of ultracold molecules'' held in February 2022. We thank John Bohn, Simon Cornish, James Croft, Philip D Gregory, Gerrit Groenenboom, Junyu He, Tijs Karman, Yi-Xiang Liu, and Kai Voges for helpful discussions and comments on the manuscript. We thank Kang-Kuen Ni's and Simon Cornish's groups for providing the graphics shown in Figs.\ \ref{fig:ni_lab_apparatus}, \ref{fig:chopped_trap_gregory}, and \ref{fig:chopped_trap_gregory_2}.  We gratefully acknowledge support from the Max Planck Society, the European Union (PASQuanS Grant No.\ 817482) and the Deutsche Forschungsgemeinschaft under Germany's Excellence Strategy -- EXC-2111 -- 390814868 and under Grant No.\ FOR 2247. A.S.\ acknowledges funding from the Max Planck Harvard Research Center for Quantum Optics.

\bibliography{bibliography}

\end{document}